\documentclass[twocolumn]{aastex63}

\usepackage{multirow}
\usepackage{graphicx}
\usepackage[space]{grffile} 

\newcommand{\xmm} {{\it XMM-Newton}}
\newcommand{\chandra} {{\it Chandra}}
\newcommand{\nustar} {{\it NuSTAR}}
\newcommand{\swift} {{\it Swift}}

\newcommand{\erosita} {eROSITA}

\newcommand{\lx} {$L_{\rm{X}}$}

\newcommand{\degree}{{$^\circ$}}
\newcommand{\ergs}{\mbox{\thinspace erg\thinspace s$^{-1}$}}
\newcommand{\ergcms}{\mbox{\thinspace erg\thinspace cm$^{-2}$\thinspace s$^{-1}$}}

\newcommand{\cntrt}{counts\,s$^{-1}$}

\shorttitle{Fast X-ray transients in \nustar\ data}
\shortauthors{Brightman et al.}

\begin{document}

\title{Fast X-ray Transients in \nustar\ data}

\author{Murray Brightman}
\affiliation{Cahill Center for Astrophysics, California Institute of Technology, 1216 East California Boulevard, Pasadena, CA 91125, USA}

\author{Joahan Casta\~{n}eda Jaimes}
\affiliation{Cahill Center for Astrophysics, California Institute of Technology, 1216 East California Boulevard, Pasadena, CA 91125, USA}

\author{Daniel Stern}
\affiliation{Jet Propulsion Laboratory, California Institute of Technology, Pasadena, CA 91109, USA}

\author{Brian Grefenstette}
\affiliation{Cahill Center for Astrophysics, California Institute of Technology, 1216 East California Boulevard, Pasadena, CA 91125, USA}

\email{murray@srl.caltech.edu}

\begin{abstract}
Fast X-ray transients (FXTs) are flashes of X-rays that last for a few hundreds of seconds to a few hours. An enigmatic population of these transients that did not have a clear origin has been known for several decades, mostly found serendipitously in soft X-ray imaging observations. Recent progress in this field by {\it Einstein Probe} has found that many FXTs are associated with gamma-ray bursts and the collapse of massive stars. Motivated by this, we searched the \nustar\ archive in the harder 3--79 keV band for $\sim1000$\,s duration transients. From 204 Ms of exposure we present five candidate FXTs, four of which are spectrally hard, with power-law indices $-3<\Gamma<0$, standing them apart from FXTs discovered in the soft band. Three have potential associations with galaxies at $z=0.1-2$, implying 3--79 keV luminosities of $10^{43}$ to $10^{48}$\ergs\ and volumetric event rates of 125--2900 Gpc$^{-3}$ yr$^{-1}$. The properties of these \nustar\ FXTs most resemble low-luminosity gamma-ray bursts, and would be much more common than their higher-luminosity counterparts in this redshift range.

\end{abstract}


\section{Introduction}
Fast X-ray transients (FXTs) are X-ray flares with durations ranging from a few hundred seconds to a few hours. Known X-ray transient types with these timescales include stellar flares, X-ray binary outbursts, supernova shock breakouts and gamma-ray bursts \citep[GRBs,][]{polzin23}. However, a population of FXTs exists where the origin is still unknown, such as the X-ray flash found in a \chandra\ observation of M86 by \cite{jonker13}. The timescale and implied luminosity of the event led the authors to suggest it was due to the tidal disruption of a white dwarf by an intermediate-mass black hole, albeit a potential host galaxy at a higher redshift was found by \cite{eappachen22} which would make this scenario less likely. A similar event presented in \cite{glennie15} and CDF-S XT1 presented by \cite{bauer17} followed and it was suggested some type of GRB could explain them (see also \cite{eappachen22} and \cite{qvasquez25}). Later, systematic searches found further candidates \citep{yang19,alp20,quirola22,quirola24}. Binary neutron star mergers \citep[e.g.][]{sarin21} or a millisecond magnetar \citep{xue19,lin22,quirola24a} have also been invoked to explain these events \citep[but see][]{qvasquez25}, but no clear explanation was presented. 

One standout event was XRO 080109 serendipitously found in a \swift\ observation that was found coincident in time and position with the supernova SN 2008D. XRO 080109  lasted about 400\,s and had a peak luminosity of \lx$\sim6.1\times10^{43}$ \ergs\ and was concluded to be a supernova shock breakout \citep{soderberg08}. However, until recently the lack of real-time discovery and rapid multiwavelength follow-up has hampered the understanding of most FXTs with almost all of them being discovered in archival \chandra\ and \xmm\ observations, and years after the event \citep[e.g.][]{glennie15,alp20,deluca21,quirola22,quirola24,khan25}. 

While \erosita\ uncovered a population of quasi-periodic eruptions on these timescales \citep[e.g.][]{arcodia24}, as well as several tidal disruption events \citep{sazonov21}, the field is now being revolutionized by the {\it Einstein Probe} mission \citep[{\it EP},][]{yuan15,yuan22}. {\it EP}, which was launched in 2024, covers $\sim3600$ square degrees of the sky in the soft 0.5--4 keV X-ray band with the lobster-eye micro-pore optics of its Wide-field X-ray Telescope. Recent results indicate that some {\it EP} FXTs are associated with GRBs \citep[e.g.][]{levan25,liu25}, though some do not feature a gamma-ray counterpart \citep[e.g.][]{sun25,vandalen25}. The first EP transient associated with a GRB was EP240219a \citep{yin24}, though it had no confirmed optical counterpart. The first source to have an observed transient counterpart at longer wavelengths was EP240315a \citep{gillanders24,levan25} which was associated with a long GRB at $z=4.859$. Shortly thereafter EP240414a was detected which had an optical counterpart at $z=0.401$ that evolved into broad-lined Type Ic supernova (SN Ic-BL), those typically associated with long GRBs, while no GRB was detected \citep{vandalen25,sun25}. EP250108a is another more recent such example of a SN Ic-BL associated with an FXT, this one at $z=0.176$ \citep{srinivasaragavan25,rastinejad25}. The luminosity of these events ranged from $10^{45}$ to $10^{50}$ \ergs\ in the X-ray band, implying that GRBs of varying luminosities could explain a significant portion of the FXT population.

So far studies of FXTs have focussed on soft X-ray energies, $E<10$ keV, with \chandra, \xmm\ and {\it EP}. \nustar\ \citep{harrison13}, which was launched in June 2012 into low-Earth orbit, operates in the 3--79 keV energy band and was the first focussing hard X-ray ($>10$ keV) observatory in orbit and is currently the most sensitive X-ray telescope above 10 keV of any operating mission. The telescope consists of two co-aligned Wolter-I type optics each with a focal plane module (FPM) labelled FPMA and FPMB consisting of four Cadmium-Zinc-Telluride detectors at the end of a 10-m extendable mast. \nustar's point spread function (PSF) has a 18\arcsec\ FWHM, and a half power diameter (HPD) of 58\arcsec. The detectors have a triggered read-out and have excellent timing resolution of 2 $\mu$s relative to the on-board clock. With its unparalleled sensitivity and spatial resolution above 10 keV and field of view comparable to \chandra, \nustar\ has now amassed more than 13 years of observations which has the potential to contain a number of FXTs at higher energies. While \nustar\ observations have been searched for serendipitous detections in the field of view of targeted observation \citep{lansbury17,greenwell24} no systematic search for FXTs has been carried out to date. Motivated by recent advances in the field of FXTs with {\it EP}, we have conducted a search for FXT candidates in \nustar\ data.

This paper is organized as the following. In Section \ref{sec_srcs} we describe our search for FXT candidates in \nustar\ data and in Section \ref{sec_fxts} we present our candidates. We discuss the nature of these candidate FXTs in Section \ref{sec_nature} and present a summary of our findings and our conclusions in Section \ref{sec_sum}.

\section{Searching for FXTs in \nustar\ data}
\label{sec_srcs}

\subsection{Source detection}

We searched all \nustar\ observations from launch in June 2012 until June 2025, totaling 5235 observations with 204 Ms of exposure. The \nustar\ data were processed with {\tt HEASOFT} \citep{heasoft} and {\tt CALDB}, where the former ranged from v6.27.1 to v6.35 and the latter ranged from v20200429 to v20250317 depending on the date of the observation. We choose to search for $\sim1000$\,s duration transients to match the timescales of other FXTs discovered to date. FXTs would not typically be detected in standard long \nustar\ exposures since the source counts from a $\sim1000$\,s event would be dwarfed by background counts in a full exposure of 20 ks or more. We therefore use the {\sc heasoft} tool {\tt xselect} to split each observation into 1000\,s slices, starting with TSTART taken from the header of the cleaned events file, in steps of 500\,s such that the time slices are overlapping. We also stipulate that each time slice must have 300 counts or more in it. 

For each time slice, we create images in the 3--79 keV, 3--8 keV, 8--24 keV and 24--79 keV bands, also using {\tt xselect}. We then use {\sc ximage} to add the FPMA and FPMB images together.  Adding the images is standard procedure for serendipitous source detection in \nustar\ \citep{lansbury17,greenwell24} and has the advantage of increasing the depth of the observation. This can be reduced however if there is any offset in the FPMA and FPMB position that can have a difference of 5--10\arcsec. Source detection is done with the {\tt detect} function in {\sc ximage} for each time slice and each energy band. The {\tt detect} command locates point sources in the co-added images using a sliding-cell method. The average background intensity is estimated by dividing the image into equal boxes to estimate the local background intensity and the event distribution in each box is checked for spatial uniformity and consistency with the expected statistical distribution.

We next considered the probability threshold for distinguishing a candidate source from a background fluctuation taking into account the number of trials undertaken. The 204 Ms of exposure time resulted in $2\times10^{6}$ independent, non overlapping 1-ks time slices. The \nustar\ PSF has a 90\% HPD of $\sim1$\arcmin\ on a field of view of 13\arcmin$\times$13\arcmin. Excluding the area within 3\arcmin\ of the target, this yields $\sim100$ independent position on the detector. Summing all trials across time and detector space gives us $\sim10^8$ trials in total. We thus stipulate that the probability that the candidate source is a background fluctuation must be less than $10^{-8}$ using the PROB\_LIMIT qualifier in {\sc ximage}. This means that we would expect less than 1 false positive in our final results for each energy band considered.

\subsection{Candidate selection}
\label{sec_selection}

We then select candidate FXTs as sources that are only detected in one time slice, and that are at least 3\arcmin\ from the target position or any other detection.  This selects against persistent sources and flares from them since these would not be considered FXTs as they are known. This yields 82 candidates in the 3--79 keV band, 29 in the 3--8 keV band, 11 in the 8--24 keV band, and 32 in the 24--79 keV band. 

\begin{figure}
\begin{center}
\includegraphics[trim=0 0 0 0, width=85mm]{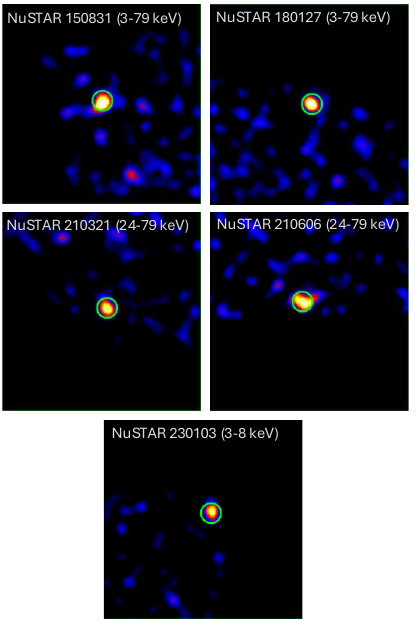}
\end{center}
\caption{\nustar\ FPMA+FPMB images of the FXT candidates, in the energy band and 1000\,s time-slice in which they were detected. The circles with 25\arcsec\ radius are centered on the detection. The images have been smoothed with a 10\arcsec\ Gaussian function.}
\label{fig_cutouts}
\end{figure}

For each of these candidates we extract source and background events  from each detector individually using {\tt nuproducts}, from a circular region with 25\arcsec\ radius for the source, and an annulus with an inner radius of 25\arcsec\ and outer radius of 120\arcsec\ for the background, both centered on the location of the candidate.  A 25\arcsec\ region was used for the source extraction in order to optimize the signal to noise ratio for these faint sources (See Figure \ref{fig_cutouts}). Scaling the background counts to the source region size, we use the Poisson likelihood function to calculate the probability that the counts detected in the source region are consistent with the background. We do this for the total counts from each telescope, yielding $P_{\rm tot}$ and separately for FPMA, yielding $P_{\rm A}$, and FPMB, yielding $P_{\rm B}$. Since FPMA and FPMB are independent detectors, we also determine the probability that a source is consistent with the background as the product of $P_{\rm A}$ and $P_{\rm B}$, i.e. $P_{\rm A,B}=P_{\rm A}\times P_{\rm B}$. We then select candidates where $P_{\rm tot}<10^{-8}$. This reduces our candidates to 11 in the 3--79 keV band, 1 in the 3--8 keV band, none in the 8--24 keV band, and 3 in the 24--79 keV band. 

We then cross-correlated the position of these candidates with the \swift\ Point Source Catalog \citep[2SXPS,][]{evans20}, the Fourth \xmm\ Serendipitous Source Catalogue \citep[4XMM,][]{webb20}, the \chandra\ Source Catalog \citep[CSC2,][]{evans10} and the the eROSITA all-sky survey catalog \citep[1eRASS][]{merloni24}, with a search radius of 30\arcsec, finding that 6 of the 3--79 keV candidates are previously known persistent X-ray sources being mostly AGN or stars. The source density in a typical 30-ks \xmm\ exposure, having a sensitivity of $10^{-15}$ \ergcms\ in the 0.5--2 keV band, is 790 deg$^{-2}$ \citep{cappelluti07}, or 0.17 X-ray sources in a 30\arcsec\ circular region. The Poisson likelihood of a chance association with one or more sources based on these numbers is 16\%. At the deepest exposures of $>1$ Ms source densities can reach 10,000 deg$^{-2}$ and the likelihood of a chance association rises 89\%. Therefore there is a chance that we miss a genuine FXT candidate by excluding those possibly associated with known X-ray sources.

After excluding known X-ray sources, we manually examine the remaining candidates, finding that 4 candidates in the 3--79 keV band and 1 candidate in the 24--79 keV band are likely spurious detections found in stray light regions or coincident with background flares. For the remaining candidates, we confirmed that the counts in the source region were consistent with the background counts during the course of the observation to check that the source region was on the detector during its entirety to ensure any motion of the telescope did not produce false variability. In the end we are left with 1 candidate in the 3--79 keV band, 1 in the 3--8 keV band and 2 in the 24--79 keV band, which we list in Table \ref{table_sources}. None of these were detected in more than one energy band. We also include in this table a candidate, NuSTAR 180127, detected in the 3--79 keV band where $P_{\rm tot}$ decreased slightly due to better background sampling, but where $P_{\rm A,B}$ is less than $10^{-8}$. This leaves us with a total of 5 candidates which we present in Table \ref{table_sources}.  \nustar\ FPMA+FPMB images of the FXT candidates, in the energy band and 1000\,s time-slice in which they were detected, are shown in Figure \ref{fig_cutouts}.

\begin{table*}
\centering
\caption{Summary of source properties}
\label{table_sources}
\begin{center}
\begin{tabular}{c c c c c c c c c c c c c c}
\hline
FXT name	& Band		& Sequence ID & \multicolumn{2}{c}{Best position} 			& Time of detection & \multicolumn{4}{c}{Counts} & \multicolumn{4}{c}{False alarm rates (log)} \\
			& (keV)		&			& RA (\degree)	& Dec (\degree)& UTC			   	&  A	& B	& Tot	& Back	& $P_{\rm A}$ & $P_{\rm B}$ & $P_{\rm tot}$ & $P_{\rm A,B}$ \\
NuSTAR 150831 & 3-79 & 60161497002 & 188.036 & 58.0032 & 2015-08-31T03:47 & 11 & 6 & 17 & 2.3 &       -7.9 &       -2.3 &       -9.3 &       -10.2\\ 
NuSTAR 180127 & 3-79 & 60375002002 & 188.001 & 14.5462 & 2018-01-27T03:16 & 6 & 11 & 17 & 3.8 &       -1.8 &       -6.9 &       -6.3 &       -8.8\\
NuSTAR 210321 & 24-79 & 90701312002 & 351.334 & 61.0808 & 2021-03-21T06:26 & 7 & 5 & 12 & 1.4 &       -5.1 &       -3.5 &       -7.6 &       -8.6\\
NuSTAR 210606 & 24-79 & 90701319002 & 238.830 & -54.2044 & 2021-06-06T05:57 & 9 & 6 & 15 & 2.3 &       -4.9 &       -3.5 &       -7.6 &       -8.3\\
NuSTAR 230103 & 3-8 & 50801001002 & 184.669 & 14.5255 & 2023-01-03T05:44 & 5 & 4 & 9 & 0.5 &       -5.1 &       -4.4 &       -8.3 &       -9.5\\ 
\hline
\end{tabular}
\tablecomments{The assumed positional uncertainty for all sources is 10\arcsec. The time of detection is the time of the beginning of the 1-ks time slice in which the source was detected. Counts are given in the band the source was detected in, for FPMA, FPMB, total and background. False alarm rates are given as logarithms for FPMA and FPMB separately, added together, and when considered as independent detectors.}
\end{center}
\end{table*}

\subsection{Lightcurve analysis}

\begin{figure}
\begin{center}
\includegraphics[trim=0 0 0 0, width=45mm]{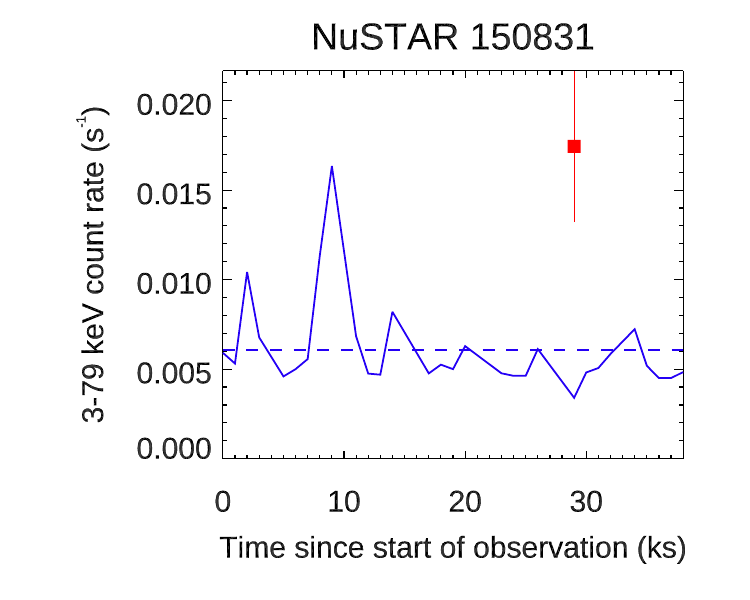}\includegraphics[trim=0 0 0 0, width=45mm]{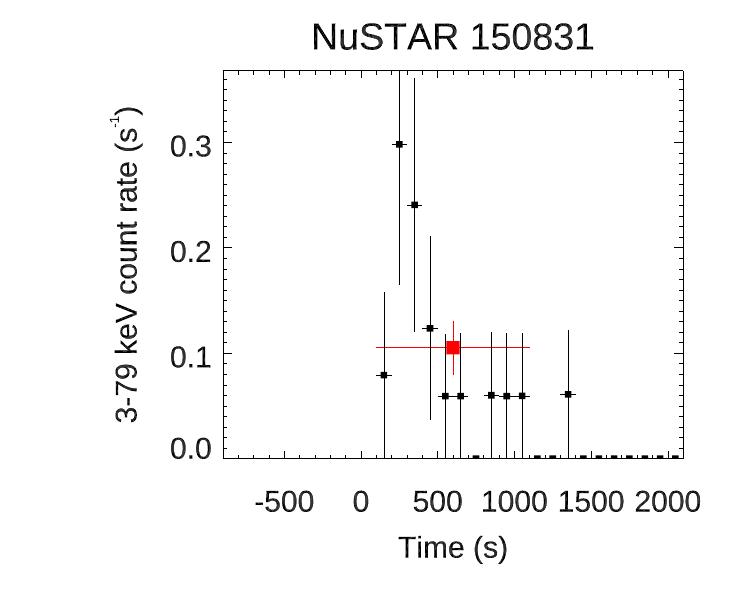}
\includegraphics[trim=0 0 0 0, width=45mm]{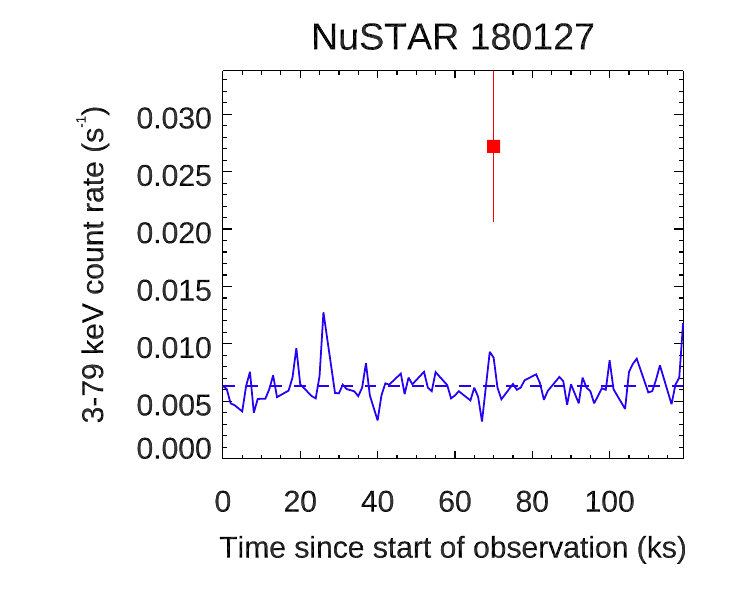}\includegraphics[trim=0 0 0 0, width=45mm]{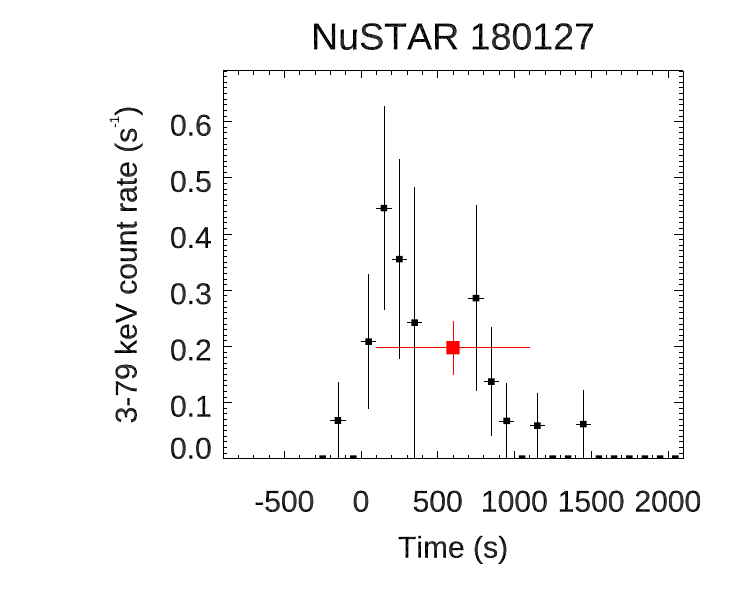}
\includegraphics[trim=0 0 0 0, width=45mm]{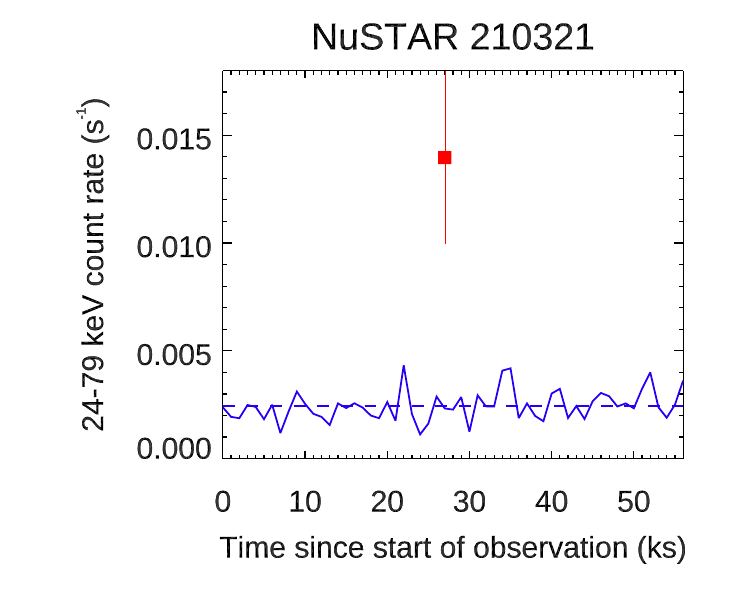}\includegraphics[trim=0 0 0 0, width=45mm]{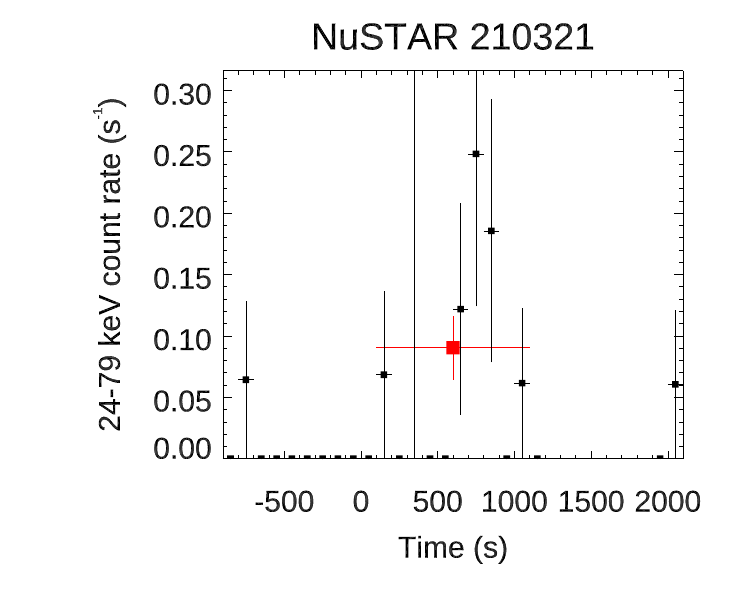}
\includegraphics[trim=0 0 0 0, width=45mm]{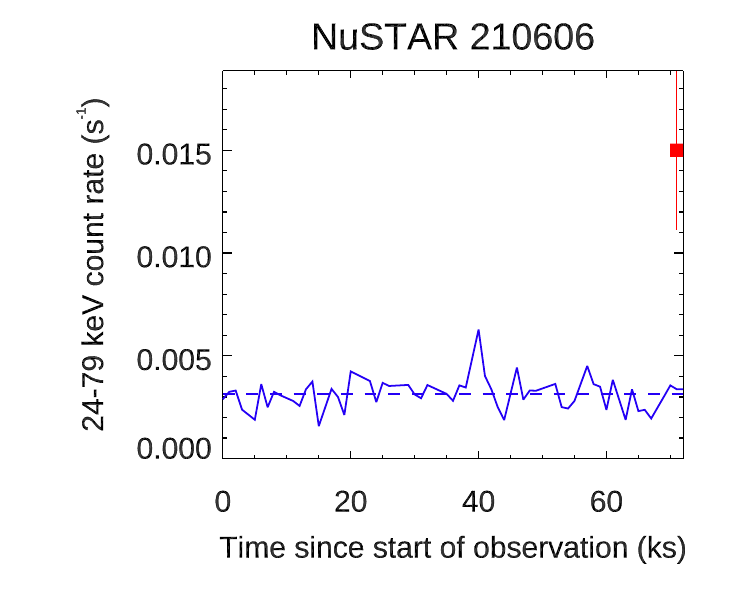}\includegraphics[trim=0 0 0 0, width=45mm]{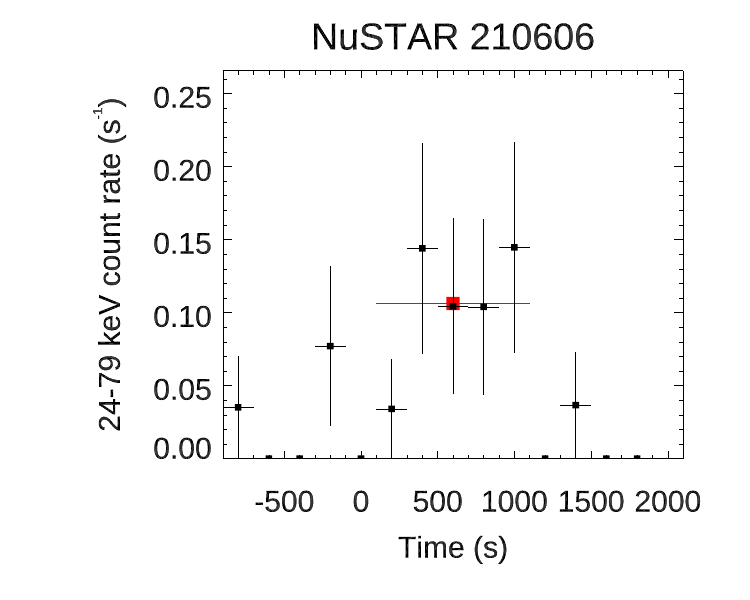}
\includegraphics[trim=0 0 0 0, width=45mm]{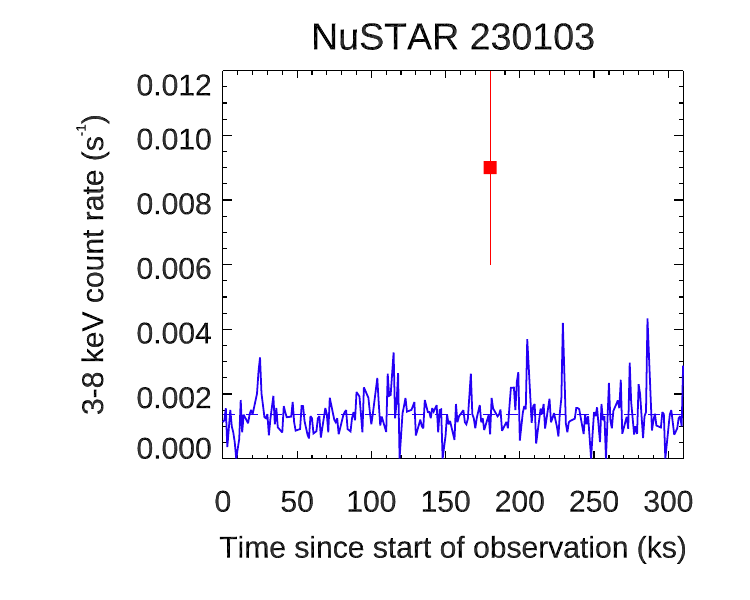}\includegraphics[trim=0 0 0 0, width=45mm]{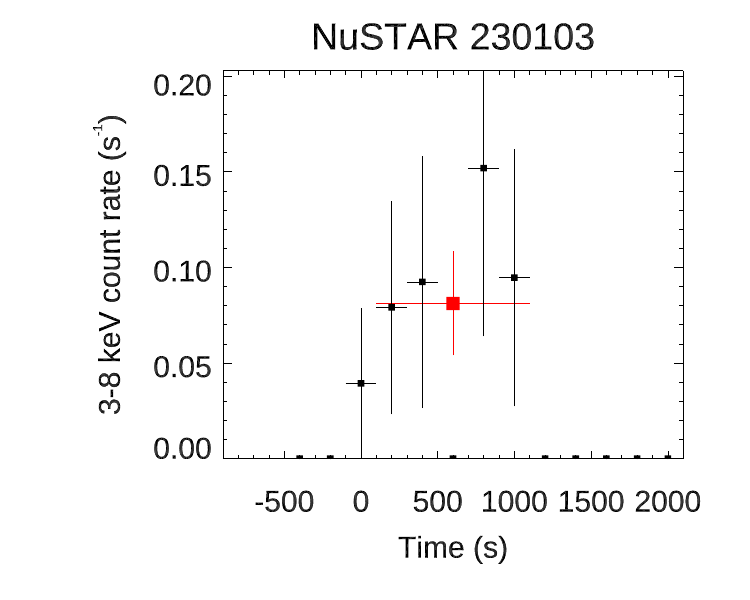}
\end{center}
\caption{Left panels - Lightcurves of the observed total count rate in the source region of the candidate \nustar\ FXTs (red data point) relative to the start time of the observation it was detected in. The scaled count rate in the background region is shown with a solid blue line (1000\,s bins) and the average background count rate is shown with a dotted blue line. Right panels - Lightcurves of the candidate \nustar\ FXTs corrected for live time, the PSF, exposure and vignetting in 100--200\,s bins (black data points) relative to the 1000\,s detection (red data point).}
\label{fig1}
\end{figure}

We extracted the lightcurves of the FXT candidates with {\tt nuproducts}, using a circular region with 25\arcsec\ radius for the source, and circular region with 120\arcsec\ radius on the same detector chip as the source for the background. We present the lightcurves of the FXT candidates in Figure \ref{fig1} which shows the observed total count rate in the source region, not corrected for live time, the PSF, exposure or vignetting. The lightcurve is binned at 1000\,s, where the FXT single detection is shown as a red square relative to the scaled background plotted with a blue line. This shows that the background is quiescent at the time of the 5 detections. The average observed total count rate for the FXTs was 9--17 counts ks$^{-1}$. We also show the lightcurve of the FXT candidates in Figure \ref{fig1} corrected for live time, the PSF, exposure and vignetting. We bin these lightcurves in 100--200\,s bins to investigate the evolution of the transient, determine the duration and look for any structure, such as a rise and fall. These corrected lightcurves show the FXTs have peak count rates of 0.15--0.45 \cntrt, and that the duration of the transients appears to be $\sim200-1000$\,s, with some evidence for a rise and fall for NuSTAR 150831, NuSTAR 180127 and NuSTAR 210321, and the possibility of a double peak structure for NuSTAR 230103. The limited number of counts in each lightcurve however, does not allow for a more detailed analysis.

\subsection{Background analysis}

 The \nustar\ background is composed of components that are both static and those that vary with time. At soft X-ray energies ($E<15$ keV) the \nustar\ background is dominated by the Cosmic X-ray Background \citep{giacconi62,gilli07}, which does not have a time variable component. There are contributions at these energies during periods of solar activity \citep{grefenstette22}, but we checked for solar flares using the GOES satellites and confirmed that none occurred during these periods. At high energies, the background is dominated by the flux of charged particles through the observatory \citep{grefenstette22}. This has a time variable component related to the location of the observatory with the South Atlantic Anomaly (SAA), the tentacle, and with the local geomagnetic rigidity cutoff. To check for this we examined the source and background rates of the candidates as a function of the \nustar\ orbit. NuSTAR 150831 and NuSTAR 180127 occurred away from both regions. NuSTAR 210321 occurred close to the tentacle region and NuSTAR 210606 occurred over the tentacle region but the background does not show any elevated count rates around this region in the observations. NuSTAR 230103 occurred close to the SAA but again the background showed no increase near this region during this observation.

\subsection{Spectral analysis}

\begin{figure*}
\begin{center}
\includegraphics[trim=0 0 0 0, width=65mm]{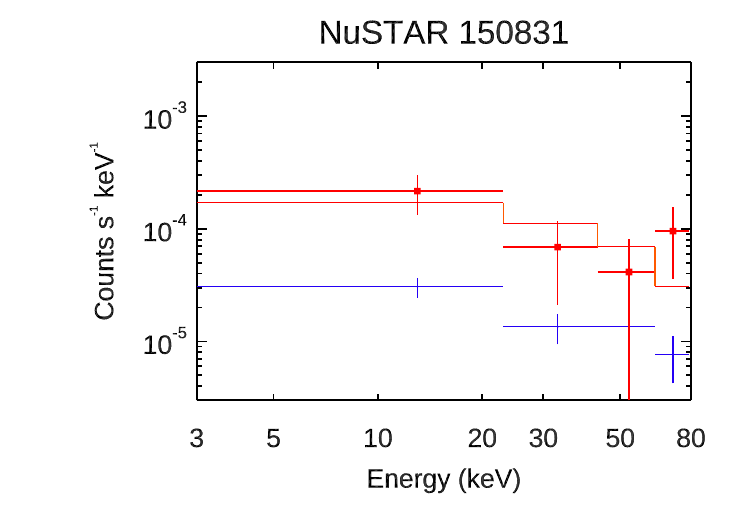}
\includegraphics[trim=0 0 0 0, width=65mm]{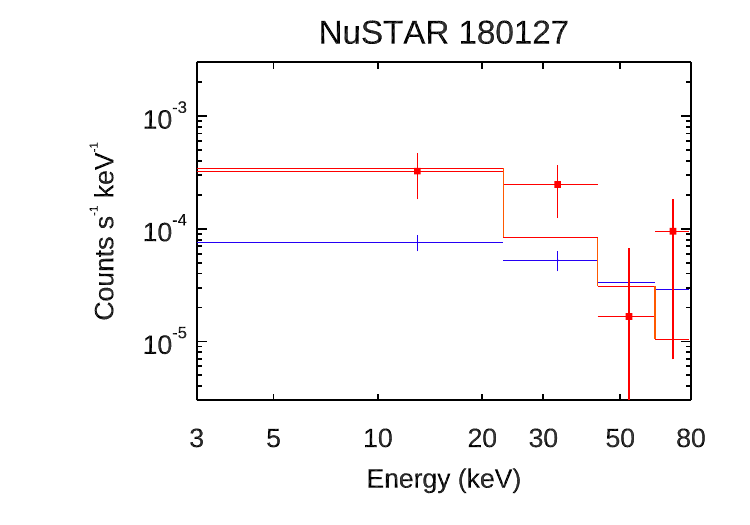}
\includegraphics[trim=0 0 0 0, width=65mm]{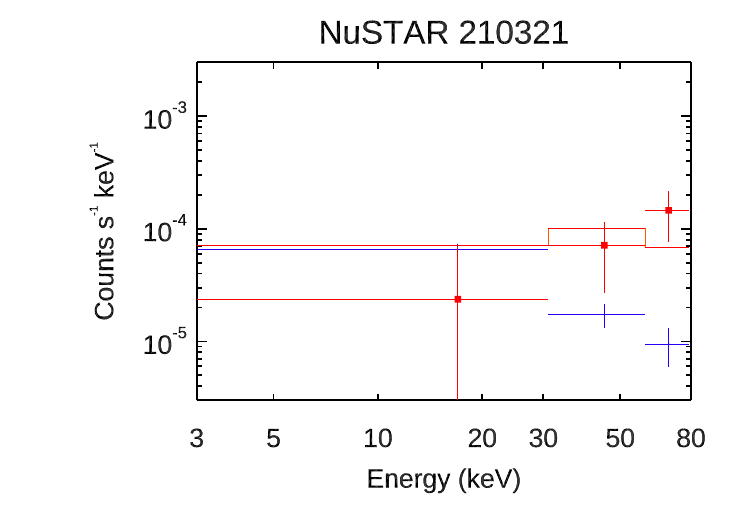}
\includegraphics[trim=0 0 0 0, width=65mm]{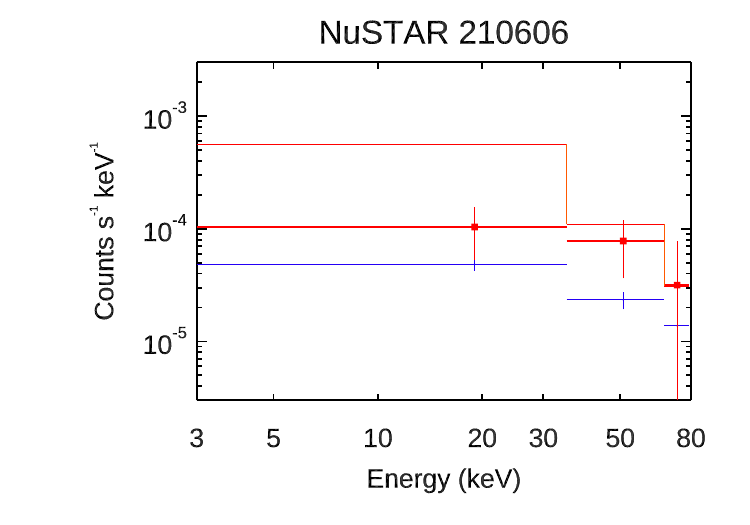}
\includegraphics[trim=0 0 0 0, width=65mm]{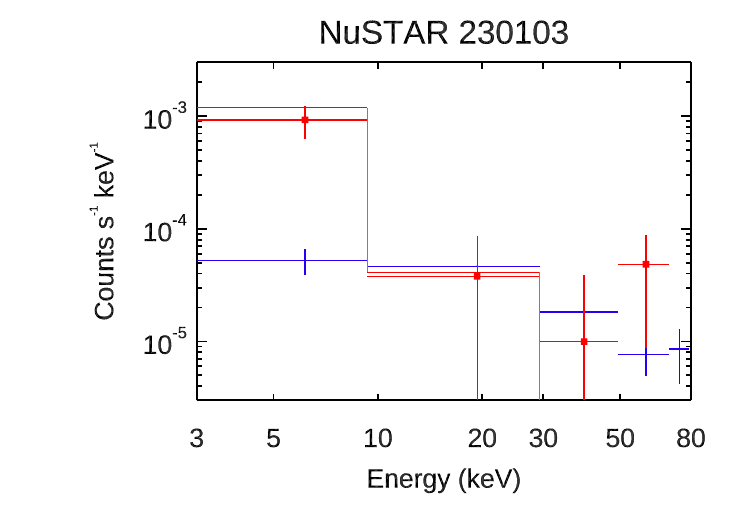}
\end{center}
\caption{Spectra of our \nustar\ FXTs where the source is shown in red and the background is shown in blue. The fitted {\tt powerlaw} model is shown as a solid red line. The FPMA and FPMB data have been grouped together and binned for plotting purposes, but were fitted separately with minimal binning.}
\label{fig3}
\end{figure*}

As with the lightcurves, we extracted the spectra of the FXT candidates with {\tt nuproducts}, using a circular region with 25\arcsec\ radius for the source, and circular region with 120\arcsec\ radius on the same detector chip as the source for the background. This circular region for the background was used instead of the annulus in Section \ref{sec_selection} since it is larger and could sample the background better. We supply a GTI file to select the 1000\,s period where the source was detected. We group the spectra with a minimum of 1 count per bin with the {\sc heasoft} tool {\tt grppha}. We then used {\sc xspec} to fit the spectra in the 3--79 keV band with the {\tt powerlaw} model, where the photon flux, $F_{\gamma}$, is described as $NE^{-\Gamma}$, where $N$ is the normalization in units of photons keV$^{-1}$ cm$^{-2}$ s$^{-1}$ at 1 keV, $E$ is the photon energy, and $\Gamma$ is the power-law index. We use the Cash statistic \citep{cash79} as the fit statistic and the background subtracted. Since the C-statistic cannot formally be used when the background is subtracted, {\sc xspec} uses a modified version of the C-statistic known as the W-statistic to account for this. We also trialled a {\tt cutoffpl} model but this did not improve the fit in any of the sources. The spectral parameters are listed in Table \ref{table_specfits} and the spectral fits are shown in Figure \ref{fig3} which show that the majority of our FXT candidates have hard spectra with power-law indices $-3<\Gamma<0$, albeit in some cases the uncertainties are consistent with softer values too.

\begin{table}
\centering
\caption{Spectral properties}
\label{table_specfits}
\begin{center}
\begin{tabular}{c c c c c c c c c c c c c c}
\hline
FXT name	& $\Gamma$		& log$F_{\rm 3-79 keV}$ & $C$ & DoFs  \\
 & & (\ergcms) & & \\
NuSTAR 150831 & $-0.8^{+0.8}_{-0.9}$ & $-10.3^{+0.4}_{-0.3}$ & 13.2 & 13 \\ 
NuSTAR 180127 & $0.2^{+5.1}_{-1.5}$ & $-10.6^{+0.6}_{-1.4}$ & 24.2 & 13 \\ 
NuSTAR 210321 & $-2.7^{+1.8}_{-3.3}$ & $-9.9^{+0.3}_{-0.4}$ & 14.6 & 9 \\ 
NuSTAR 210606 & $0.4^{+2.3}_{-1.8}$ & $-10.1^{+0.3}_{-0.3}$ & 15.2 & 11 \\ 
NuSTAR 230103 & $3.2^{+2.0}_{-1.5}$ & $-11.5^{+0.3}_{-0.3}$ & 18.2 & 11 \\ 
\hline
\end{tabular}
\end{center}
\end{table}

\subsection{Positional analysis}

In order to determine the best positions of our sources, we attempt to use the position of the target of the observation relative to its catalogued position to correct for any astrometric error. We do this for each time slice since the source can move around slightly on the detector and take the mean of these positions. This is only possible for NuSTAR 150831 and NuSTAR 210606 since for the other sources the target of the observation is too faint to be detected in the individual time slices. We list the results in Table \ref{table_positions} giving the name of the target, its mean position, its catalogued position from SIMBAD, the mean offsets, the observed source position and the corrected source position. We find for these two sources that the average offset was less than 5\arcsec. However, once this offset has been accounted for, the statistical uncertainty on the position remains, which is a few arcseconds. Analysis from the \nustar\ serendipitous source survey found the average astrometric offset between a \nustar\ source and its soft X-ray counterpart from CSC2, 4XMM or 2SXPS was $\sim10$\arcsec\ \citep{lansbury17,greenwell24}. To be consistent and conservative, we adopt this as our positional uncertainty for all sources. We show the position of our candidate FXTs relative to known catalogued sources in Figure \ref{fig_counterparts}.

\begin{table*}
\centering
\caption{Source positions}
\label{table_positions}
\begin{center}
\begin{tabular}{c c c c c c c c c c c c c c}
\hline
FXT & Target name		& \multicolumn{4}{c}{Target position} & \multicolumn{1}{c}{Offsets} & \multicolumn{4}{c}{Source position} \\
& &  \multicolumn{2}{c}{Measured} & \multicolumn{2}{c}{Catalogued} &  & \multicolumn{2}{c}{Measured} & \multicolumn{2}{c}{Corrected}  \\
			&			& RA	 (\degree)	& Dec (\degree)& RA (\degree)	& Dec (\degree)& RA/Dec/R(\arcsec) & RA (\degree)	& Dec (\degree) & RA (\degree) & Dec (\degree) \\
NuSTAR 150831 & NGC 4500 	& 187.8417 & 57.9643 	& 187.8424 & 57.9646		& 2.4/1.3/4.4	 & 188.0354 & 58.0028	& 188.0361 & 58.0032 \\ 
NuSTAR 210606 & SGR J1555-54 & 238.7881 & -54.0618	& 238.7857 & -54.0608	& -8.7/3.6/2.3	 & 238.8321     & -54.2054	& 238.830     & -54.2044 \\ 
\hline
\end{tabular}
\end{center}
\end{table*}

\section{\nustar\ FXT candidates}
\label{sec_fxts}

\subsection{NuSTAR 150831}
NuSTAR 150831 was detected in the 3--79 keV band at 2015-08-31T03:47 at a position of RA=188.03610\degree\ Dec=58.003191\degree\  ($l=127.88588$\degree, $b=58.939245$\degree) with a false alarm rate of $5\times10^{-10}$. It was found in an observation of NGC 4500 which had an exposure of 21\,ks (sequence ID 60161497002). The spectrum was hard with a power-law index $\Gamma=-1\pm1$, a 3--79 keV flux of $6^{+6}_{-3}\times10^{-11}$ \ergcms\ and a flux of $2^{+3}_{-1}\times10^{-13}$ \ergcms\ when extrapolated into the 0.3--10 keV band.  

No X-ray source has been detected at the location of this source despite five \swift\ observations of the source region which took place prior to the time of the transient, four several years beforehand, one only a few hours previous at 2015-08-30T21:09. The stacked $3\sigma$ count rate upper limit for these observations is $6.6\times10^{-4}$ \cntrt\ as determined by the online tool provided by the University of Leicester\footnote{https://www.swift.ac.uk/user\_objects/} \citep{evans07,evans09}. This count rate corresponds to a 0.3--10 keV flux of $1.4\times10^{-13}$ \ergcms\ assuming a photon index of $\Gamma=-1$. \chandra\ and \xmm\ have not observed the source region.

While two GRBs were detected by \swift/BAT on 2015-08-31, GRB 150831A \citep{lien15} and GRB 150831B \citep{marshall15}, neither were positionally or temporally coincident with NuSTAR 150831. 

\begin{figure*}
\begin{center}
\includegraphics[trim=0 0 0 0, width=160mm]{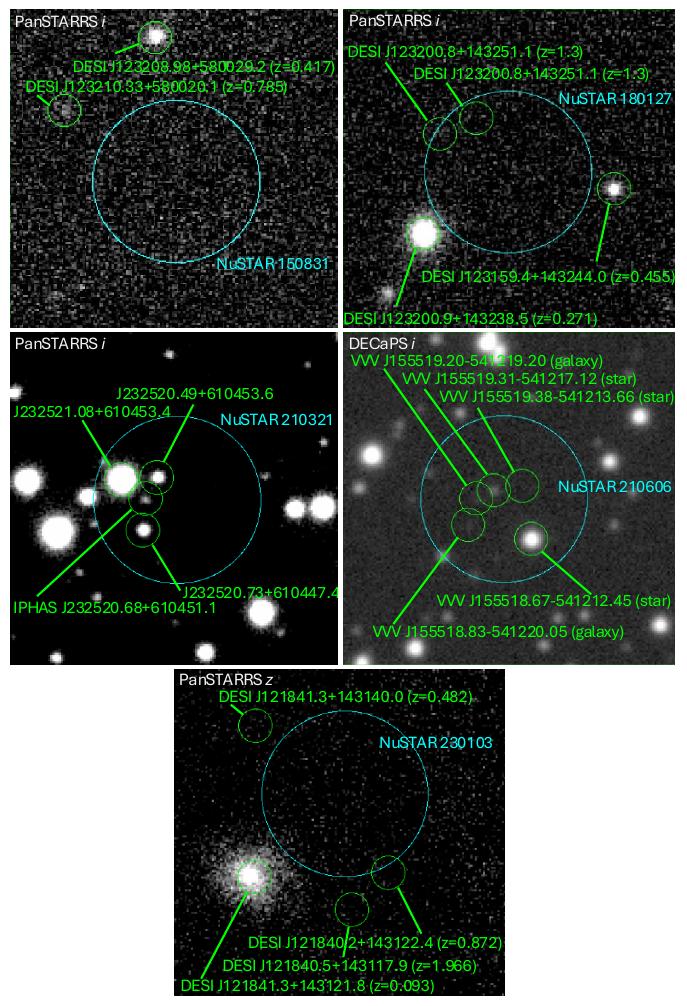}
\end{center}
\caption{PanSTARRS $i$-band images of the regions near NuSTAR 150831, NuSTAR 180127 and NuSTAR 210321, DECaPS $i$-band imaging of NuSTAR 210606 and PanSTARRS $z$-band imaging of NuSTAR 230103. The cyan circles, which have a radius of 10\arcsec, represent the positional uncertainties of the FXTs. The potential counterparts have been labelled.}
\label{fig_counterparts}
\end{figure*}

NuSTAR 150831 is 6.7\arcmin\ offset from the center of the spiral galaxy NGC 4500 which has mean redshift-independent distance of 49 Mpc taken from NED. At this distance the projected separation is 95 kpc, and the implied luminosity is $2^{+2}_{-1}\times10^{43}$\ergs. There are also two background galaxies close to the \nustar\ source, DESI J123210.33+580020.1 which is 16\arcsec\ away and DESI J123208.98+580029.2 which is 18\arcsec\ away  which we show in Figure \ref{fig_counterparts}. These offsets are larger than our adopted 10\arcsec\ positional uncertainty, but the \nustar\ positional uncertainty can be this large for faint sources \citep{greenwell24}. The galaxies have photometric redshifts, $z_{\rm phot}$, of 0.785$\pm0.654$ and 0.417$\pm0.034$ respectively \citep{duncan22}. The sky density of DESI sources in this region of the sky is 12 arcmin$^{-2}$. Therefore the probability of finding one of these galaxies within these offsets by chance is $>99$\% based on the Poisson likelihood function. However, if NuSTAR 150831 is associated with one of the DESI galaxies, the implied 3--79 keV luminosity would be either $2^{+2}_{-1}\times10^{47}$ or $4^{+4}_{-2}\times10^{46}$\ergs\ depending on which galaxy is assumed.

\subsection{NuSTAR 180127}
NuSTAR 180127 was detected in the 3--79 keV band at 2018-01-27T03:16 at a position of RA=188.001\degree\ Dec=14.5462\degree\  ($l=282.171$\degree, $b=76.6281$\degree) with a false alarm rate of $5\times10^{-7}$. It was found in an observation of NGC 4501 which had an exposure of 63\,ks (sequence ID 60375002002). The spectrum was hard with a power-law index $\Gamma=0^{+5}_{-1}$, a 3--79 keV flux of $3^{+8}_{-2.9}\times10^{-11}$ \ergcms\ and a flux of $6^{+165}_{-5}\times10^{-13}$ \ergcms\ when extrapolated into the 0.3--10 keV band.  

No GRBs were reported on 2018-01-27 and no X-ray source has been detected at the location of this source despite 4 \swift\ observations prior to the occurrence of the transient and 4 afterwards. One \swift\ observation took place only a few hours before at 2018-01-26T11:47, and another only a few months afterwards. \chandra\ observed the source region three times in 2002 with a total upper limit of $5\times10^{-15}$ \ergcms\ and \xmm\ observed once in 2001 with an upper limit of $1\times10^{-15}$ \ergcms. A second \nustar\ observation of NGC 4501 occurred on 2018-05-21 but NuSTAR 180127 was not in the field of view.

NuSTAR 180127 is 7.6\arcmin\ offset from the center of the spiral galaxy NGC 4501 which has mean redshift-independent distance of 18 Mpc. At this distance the projected separation is 40 kpc, and the implied luminosity is $1.09^{+3}_{-1.05}\times10^{42}$\ergs. There are two DESI DR8 galaxies within our adopted positional accuracy of 10\arcsec\ from NuSTAR 180127, DESI J123200.5+143253.0 with a separation of 8\arcsec\ at $z_{\rm phot}=1.1\pm0.3$ and DESI J123200.8+143251.1 with a separation of 10\arcsec\ at $z_{\rm phot}=1.3\pm0.3$ (Figure \ref{fig_counterparts}).  Two other lower-redshift galaxies lie just outside the 10\arcsec\ radius. The probability of finding a DESI source in this region of the sky by chance within a 10\arcsec\ region is 90\%. If NuSTAR 180127 is associated with one of these $z\sim1$ galaxies, the implied 3--79 keV luminosity would be either $2^{+6}_{-1.9}\times10^{47}$ or $3^{+9}_{-2.9}\times10^{47}$\ergs.

\subsection{NuSTAR 210321}
NuSTAR 210321 was detected in the 24--79 keV band at 2021-03-21T06:26 at a position of RA=351.334\degree\ Dec=61.0808\degree\  ($l=112.712$\degree, $b=-0.0706$\degree) with a false alarm rate of $3\times10^{-8}$. It was found in an observation of PNV J23244760+6111140 which had an exposure of 41\,ks (sequence ID 90701312002). The spectrum was hard with a power-law index $\Gamma=-3^{+2}_{-3}$, a 3--79 keV flux of $1.2\pm0.8\times10^{-10}$ \ergcms\ and a flux of $1^{+13}_{-0.6}\times10^{-14}$ \ergcms\ when extrapolated into the 0.3--10 keV band.  

GRB 210321A was reported on 2021-03-21T07:22:33 but not positionally coincident with NuSTAR 210321 \citep{dichiara21} and no X-ray source has been detected at the location of this source despite 12 \swift\ observations prior to the occurrence of the transient and 72 afterwards. One \swift\ observation took place a few weeks before at 2021-01-20T19:10, and another only a few days afterwards 2021-03-24T01:24:13. \chandra\ and \xmm\ have to date not observed the source region.

NuSTAR 210321 is positionally coincident with the plane of our Galaxy with 3 Gaia DR3 stars with distances of 1--7 kpc catalogued within 10\arcsec\ \citep[][Figure \ref{fig_counterparts}]{gaia23}. To verify that these three sources are indeed Galactic and search for potential peculiarities, we obtained spectra of all three sources with the Next Generation Palomar Spectrogram on the Hale 200-inch Telescope at Palomar Observatory.  This is a long-slit optical spectrograph that currently covers 6000 \AA\ to 1 $\mu$m. Upgrades over the coming months will extend this spectral coverage across the full optical window.  The data were obtained on UT 2025 August 24, which had average seeing of 1.5\arcsec\ and intermittent thin cirrus. We obtained a 300~s observation of the brightest source, J232521.08+610453.4 ($r=16.6$), and two 900~s observations with the slit aligned to simultaneously observe the two fainter sources, J232520.73+610447.4 ($r=19.6$) and J232520.49+610453.6 ($r=20.2$).  The spectra reveal all three sources to be Galactic, with absorption from H$\alpha$ and the Calcium triplet.  A fourth source, IPHAS J232520.68+610451.1 ($r=21.2$), with unknown classification, also lies within 10\arcsec\ \citep{barentsen14}. Given the sky density of Gaia DR3 sources in the region of 29 arcmin$^{-2}$, the probability of finding a Gaia DR3 source by chance within a 10\arcsec\ radius circular region is 92\%. If NuSTAR 150831 is associated with one of the three stars, the implied 3--79 keV luminosity would be between $1.5\pm0.9\times10^{34}$ and $7\pm5\times10^{35}$\ergs. \cite{gaia23} lists the luminosity of the brightest star mentioned above as 1.4 $L_{\odot}$ ($5.3\times10^{33}$ \ergs), several times less than the X-ray luminosity if NuSTAR 150831 is associated with it. This is an unlikely scenario considering the hard X-ray contribution to stellar flares is only a few percent \citep[e.g.][]{hu25}.

\subsection{NuSTAR 210606}
NuSTAR 210606 was detected in the 24--79 keV band at 2021-06-06T05:57 at a position of RA=238.830\degree\ Dec=-54.2044\degree\  ($l=327.801$\degree, $b=-0.4615$\degree) with a false alarm rate of $3\times10^{-8}$. It was found in an observation of SGR J1555.2-5402 which had an exposure of 38\,ks (sequence ID 90701319002). The spectrum was hard with a power-law index $\Gamma=0^{+2}_{-1}$, a 3--79 keV flux of $8^{+9}_{-4}\times10^{-11}$ \ergcms\ and a flux of $3.8^{+107}_{-3.6}\times10^{-12}$ \ergcms\ when extrapolated into the 0.3--10 keV band.  

GRB 210606A was detected by Fermi/GBM on 2021-06-06T03:56:02 but not positionally coincident with NuSTAR 210606 \citep{grb210606a} and no X-ray source has been detected at the location of this source despite 19 \swift\ observations prior to the occurrence of the transient and 89 afterwards. One \swift\ observation took place a few hours before at 2021-06-05T10:37, and another around a day afterwards at 2021-06-07T12:02. \chandra\ and \xmm\ have to date not observed the source region.

Like NuSTAR 210321, NuSTAR 210606 is positionally coincident with the plane of our Galaxy. Within 10\arcsec\ are 5 sources catalogued by the VISTA Variable in the Via Lactea Survey (VVV) DR4.2, two of which are likely galaxies, and three are stars which have Gaia DR3 distances of 4--5 kpc \citep[Figure \ref{fig_counterparts},][]{saito12,minniti23}. The MIR source CatWISE J155518.89-541220.4 is coincident with one of the likely galaxies and has W1$>$17.3 and W2=14.12$\pm0.04$ \citep{marocco21}. The W1-W2 color of at least 3.2 is much larger than the W1-W2$>$0.8 criterion that suggests the presence of hot dust heated by an AGN \citep{stern12,assef13}. The probability of finding a Gaia DR3 source within 10\arcsec\ in this region of the sky is 97\%. If NuSTAR 210606 is associated with one of the stars, the implied 3--79 keV luminosity would be $\sim2\times10^{35}$\ergs. 

\subsection{NuSTAR 230103}
NuSTAR 230103 was detected in the 3--8 keV band at 2023-01-03T05:44 at a position of RA=184.669\degree\ Dec=14.5255\degree\  ($l=270.095$\degree, $b=75.2662$\degree) with a false alarm rate of $5\times10^{-9}$. It was found in an observation of NGC 4254 X2 which had an exposure of 152\,ks (sequence ID 50801001002). The spectrum was soft with a power-law index $\Gamma=3^{+2}_{-1}$, a 3--79 keV flux of $3^{+4}_{-2}\times10^{-12}$ \ergcms\ and a flux of $6^{+2}_{-4}\times10^{-12}$ \ergcms\ when extrapolated into the 0.3--10 keV band.  

No GRBs were reported on 2023-01-03 and no X-ray source has been detected at the location of this source despite 32 \swift\ observations, all but one of which were taken before the time of the transient. \chandra\ has observed the source region six times, three times prior to the transient, the last one in 2021 and three times afterwards, two of which occurred a few months afterwards on 2023-03-31. These observations had a sensitivity of $3\times10^{-15}$ \ergcms. \xmm\ observed the source region twice only two days prior to the transient on 2023-01-01, with two other observations a few years before that. These had a sensitivity of $2\times10^{-15}$ \ergcms.

NuSTAR 230103 is 6.9\arcmin\ offset from the center of the spiral galaxy NGC 4254 which has mean redshift-independent distance of 15 Mpc. At this distance the projected separation is 31 kpc, and the implied luminosity is $8^{+8}_{-4}\times10^{40}$\ergs. There are four DESI DR8 galaxies catalogued at 10, 13, 14 and 15\arcsec\ from the position of NuSTAR 180127 with $z_{\rm phot}=0.872\pm0.459$, 0.482$\pm0.354$, 1.966$\pm0.626$, and 0.093$\pm0.021$ which are shown in Figure \ref{fig_counterparts}. A mid-infrared source WISE J121840.74+143131.1 lies only 3\arcsec\ away with  W1=18.17$\pm$0.12 and W2=17.17$\pm$0.20 where the color indicates the possibility of an AGN. The probability of finding a DESI source within 10\arcsec\ by chance in this region of the sky is 88\%. However, low redshift galaxies are rarer, such as the one at $z=0.093$. The sky density of $z<0.1$ galaxies in this region of the sky is 0.12 arcmin$^{-2}$ and the probability of one of these sources lying within 15\arcsec\ by chance is 2.3\%, much lower than the other probabilities of chance associations making this a strong candidate host galaxy for NuSTAR 230103. The implied 3--79 keV luminosity for a source at $z=0.09$ would be $7^{+8}_{-3}\times10^{43}$ \ergs, while for a source at $z=1.966$ it would be $9^{+11}_{-4}\times10^{46}$ \ergs.

\section{The nature of the \nustar\ FXT candidates}
\label{sec_nature}

\subsection{A search for transient counterparts at other wavelengths}

In our first step in determining the nature of our FXT candidates, we search for potential transient counterparts at other wavelengths. \swift\ observed a few of our FXT candidates close in time to the event. While no counterpart source was detected in the XRT we checked the UVOT images for any optical or UV counterpart. For NuSTAR 180127 and NuSTAR 210606 multiple \swift\ observations took place close in time to the X-ray trigger of the transient. We examined the UVOT images of these observations but no source was evident within 25\arcsec\ of the \nustar\ source position. For NuSTAR 210321 which has 84 \swift\ observations of the source region with some close in time to the transient, 1 UVOT source is seen nearby, mainly in the UVW1 filter, but it shows no obvious variability. For NuSTAR 150831 all \swift\ observations took place prior to the occurrence of the transient and for NuSTAR 210606 all but one did. The typical sensitivity of these observations is 21--22 mag in the UV bands.

We searched the transient name server\footnote{https://www.wis-tns.org/} (TNS) but did not find any transients catalogued within 30\arcsec\ of the \nustar\ positions which includes a catalog of FRBs. We similarly searched the ZTF alert stream using the Automatic Learning for the Rapid Classification of Events (ALeRCE) explorer\footnote{https://alerce.online/}, the ASAS-SN Sky Patrol Photometry Database\footnote{http://asas-sn.ifa.hawaii.edu/skypatrol/}, and acquired forced photometry from the ATLAS forced photometry server\footnote{https://fallingstar-data.com/forcedphot/} and Pan-STARRS\footnote{https://catalogs.mast.stsci.edu/panstarrs/} finding no candidate transients spatially and temporally coincident with the \nustar\ ones.

The lack of a transient counterpart at other wavelengths does not rule out any source type, however.  Ground-based wide-field surveys such as ZTF probe optical transients such as core-collapse supernovae out to $z<0.05$ and SN Ia out to $z<0.1$ \citep{perley20}, whereas our FXTs appear to be at higher redshifts.  Likewise, the non-detection of an optical/UV counterpart in Swift/UVOT data also does not rule out a supernova origin since most supernovae detected with UVOT are relatively nearby \citep[$z<$ 0.02,][]{brown09}. Supernovae at high redshifts have mostly been observed with HST \citep{riess04} and unfortunately HST has not observed any of the source regions of our FXTs. The soonest \swift\ observation after a transient candidate was NuSTAR 210606 which \swift\ observed 30 hours ($\sim10^{5}$\,s) afterwards. At this time, the optical/UV flux of GRB afterglows has faded significantly and in many cases not detected by the UVOT. Furthermore, only 52\% of long GRBs are detected in the UVOT when observed within the first 500\,s \citep{roming17,oates23}. Therefore the non-detection of our transient candidates by UVOT does not preclude their GRB nature.

\subsection{Volumetric event rate estimate}

NuSTAR 150831, NuSTAR 180127 and NuSTAR 230103 are potentially associated with local galaxies at distances of 49, 18 and 15 Mpc respectively. The average volume that encloses these galaxies is 8.4$\times10^4$ Mpc$^3$. These 3 candidates were found in 204 Ms of exposure time yielding a detection rate of 0.5 yr$^{-1}$. \nustar's 13\arcmin$\times$13\arcmin\ field of view covers $10^{-6}$ of the 40000 deg$^2$ sky which implies a full sky rate of $5\times10^{5}$ yr$^{-1}$. Given the volume estimate above this implies a volumetric rate of 6 Mpc$^{-3}$ yr$^{-1}$. For comparison, core-collapse supernovae have a local volumetric event rate of $7\times10^{-5}$ Mpc$^{-3}$ yr$^{-1}$ \citep{pessi25}. If our FXTs are associated with the nearby galaxies, they would be far too numerous to be associated with core-collapse supernovae.
 
Alternatively, we can estimate a volumetric event rate for these sources assuming that these events are associated with one of the background galaxies found nearby. We consider the source with the closest projected separation as the most likely host galaxy. These have redshifts of 0.417, 1.1 and 0.09 and distances of 2.4, 7.6 and 0.42 Gpc which averages out to 3.5 Gpc in an average volume of 175 Gpc$^3$. Given the full sky rate of $5\times10^{5}$ yr$^{-1}$ this implies a volumetric rate of 2900 Gpc$^{-3}$ yr$^{-1}$ with large uncertainties due to small number statistics and host galaxy identification. 

The furthest galaxy candidates are at $z=0.785$ (4.9 Gpc), $z=1.3$ (9.1 Gpc) and $z=2.0$ (16 Gpc) with an average distance of 9.9 Gpc in an average volume of 4000 Gpc$^3$. This implies the range in the volumetric rate from the uncertain host galaxy identification is 125--2900 Gpc$^{-3}$ yr$^{-1}$. The core-collapse supernova event rate rises to $9.5\times10^{5}$ Gpc$^{-3}$ yr$^{-1}$ at $z=1.1$ \citep{dahlen12} however rarer types of supernovae such as Type Ic-BL, associated with the afterglow of long-duration GRBs, have a local volumetric event rate of 350 Gpc$^{-3}$ yr$^{-1}$ \citep{pessi25}. This would be consistent with the volumetric event rate of our FXTs if they were associated with the high-redshift background galaxies, even if they had an order of magnitude higher rates at $z=1$.

\subsection{Stellar flares}

Two of our FXT candidates, NuSTAR 210321 and NuSTAR 210606, were coincident with the plane of our Galaxy with Gaia DR3 stars with distances 1--7 kpc catalogued within 10\arcsec. This could imply that the X-ray sources were stellar flares which are triggered by the release of magnetic energy in stellar atmospheres \citep{favata03}. While the timescales of stellar flares, being 100s of seconds to a day, matches those of our transients, several properties do not match. Firstly, the implied luminosities of our transients, being in the range $10^{34}$ to $10^{37}$\ergs\ is much higher than those seen in stellar flares, being $10^{29}$ to $10^{32}$\ergs\ \citep{pye15}. We found that the ratio of the X-ray to bolometric luminosities is also too high. Secondly, the spectra of NuSTAR 210321 and NuSTAR 210606 are very hard whereas the spectra of stellar flares are usually soft and thermal \citep[e.g.][]{hu25}. Finally, stellar flares are a repeating phenomenon whereas our transients have only been detected once, despite multiple \swift\ observations of both NuSTAR 210321 and NuSTAR 210606. We therefore find that stellar flares are not likely the origin of these transients.

\subsection{Magnetar flares}

Magnetar flares are produced by changes in the magnetic field of highly magnetized neutron stars known as magnetars. The most powerful examples of these are known as magnetar giant flares (MGFs) that release $10^{44}-10^{46}$ \ergs\ of energy in the time span of milliseconds \citep[e.g.][]{trigg25}. The X-ray spectra of these bursts have powerlaw index values $\Gamma$ that vary from between -1 and 0. While the spectral properties of MGFs are consistent with those of our FXTs, the ms-scale durations of MGF are much shorter than the 200--1000\,s duration of our FXTs. Furthermore, if NuSTAR 150831 and NuSTAR 180127 are associated with one of the galaxies near their positions, the implied luminosities range from $10^{46}$ to $10^{48}$\ergs, making them more powerful than MGFs. Furthermore, the local volumetric event rate of MGFs has been estimated to be $3.8\times10^{5}$ Gpc$^{-3}$ yr$^{-1}$ \citep{burns21}, far higher than the rate estimated for our FXTs.

On the other hand, magnetar outbursts, which are less powerful than MGFs, are characterized by an increase in X-ray luminosity by a factor of 10-1000 with respect to the quiescent X-ray emission of a magnetar with peak luminosities in the $10^{34}-10^{36}$ \ergs\ with durations of days in length \citep{rea11,czelati18}. If our FXT candidates originate in our Galaxy, their luminosities would be consistent with these, but their durations are much shorter. Further, no quiescent X-ray emission has been detected from our candidates, despite deep upper limits from \chandra\ and \xmm, as would be expected from a magnetar in our Galaxy.

\subsection{Gamma-ray bursts}

GRBs are usually split in two classes by their durations,  where short GRBs typically emit 90\% of their total power in less than 2\,s ($T_{90}<2$\,s) and long GRBs emit it in the $2-10^3$\,s range \citep{kouveliotou93}. Short GRBs are thought to mainly originate from binary neutron star mergers such as GW170817 \citep{gw170817} and long GRBs are thought to be produced by the collapse of massive stars, otherwise known as collapsars \citep{woosley06}. Our FXTs have timescales comparable to long GRBs, and furthermore GRB low-energy X-ray spectral indices have been observed to be in the range $-2<\Gamma<-1$ \citep{gruber14} as observed in our sources.

Long GRBs have typical luminosities of $10^{51}-10^{53}$ \ergs\ and therefore more powerful than our sources. However a population of low-luminosity GRBs (LLGRBs) have been found and are characterized by luminosities in the range of $10^{46}-10^{48}$ \ergs\ when assuming isotropic emission and have durations of 10\,s to an hour \citep{bromberg11}. While less energetic, LLGRBs are potentially much more common, with observed local volumetric rates of $\sim230$ Gpc$^{-3}$ yr$^{-1}$ compared to $\sim1$ Gpc$^{-3}$ yr$^{-1}$ for their higher luminosity counterparts \citep[e.g.][]{soderberg06,liang07,virgili09}. The luminosity, duration and volumetric rates of the \nustar\ FXTs are thus consistent with LLGRBs. \cite{liang07} showed the volumetric rate of LLGRBs dropping steeply above $z=$0.1, however this is likely due to the limitation of \swift/BAT at detecting higher redshift LLGRBs. If our FXT candidates are indeed LLGRBs, our results would suggest the event rates of LLGRBs remain high up to a redshift of around 1.

Since long GRBs are associated with the collapse of massive stars, they are usually found in star formation regions. NuSTAR 230103 lies 15\arcsec\ away from a potential host galaxy at $z=0.093$. The chance association with a galaxy at $z<0.1$ is 2\% thus we have identified it as a strong host galaxy candidate. The projected separation is $27\pm18$ kpc where the uncertainty is dominated by the positional accuracy of 10\arcsec.This is larger than typical long GRB-host galaxy offsets and their associations with star forming regions. Short GRBs associated with kilonovae produced by compact object mergers usually have larger offsets, with the largest offset being found for the short GRB 060502B, which had an offset of 17\arcsec\ from its $z=$0.287 host galaxy, implying a projected physical separation of 70 kpc \citep{bloom07}.

While our FXTs have timescales comparable to long GRBs, it has been shown that long GRBs can also be produced by compact object mergers that produce kilonovae \citep{rastinejad22}. The binary neutron star merger rate has been inferred to be between 10 and 1700 Gpc$^{-3}$ yr$^{-1}$ \citep{abbott23}, and is therefore consistent with the rate of our FXTs, therefore some of our FXTs may be produced by compact object mergers.

\subsection{Supernova shock breakouts}

It has been suggested that LLGRBs may be a physically distinct phenomenon, such as supernova shock breakouts (SN SBOs) \citep{colgate74, klein78,norris03,virgili09,irwin25}. SN SBOs are produced when the radiation driven shock produced inside the collapse of a massive star reaches the surface and breaks out releasing X-rays \citep{nakar10}. The only SN SBO to be observed in X-rays to date is that of SN 2008D, serendipitously observed in a \swift/XRT observation \citep{soderberg08}. The X-ray source known as XRO 080109 had $\Gamma=2.2$, lasted about 400\,s and had a peak luminosity of \lx$\sim6.1\times10^{43}$ \ergs. A search for further SBO candidates in the \xmm\ archive yielded 12 sources with \lx$=10^{43}-10^{46}$ \ergs\ and soft spectra \citep[$\Gamma=2-4$,][]{alp20}, albeit one was found to likely originate from a flare star \citep{eappachen24}. While the timescales and luminosities of our FXTs are consistent with SN SBOs, the spectral properties are possibly at odds. Most of our FXT candidates have hard spectra compared to the soft spectra of SN SBOs, albeit in some cases the uncertainties on our sources are consistent with softer values too. 

\section{Summary and Conclusions}
\label{sec_sum}

Motivated by recent advances in the field of FXTs, we have conducted a search for such events in \nustar\ data collected over the 13 year period 2012--2025. We found 5 candidate FXTs in the \nustar\ archive, two detected in the 3--79 keV band, two in the 24--79 keV band, and one in the 3--8 keV band. The transients have $\sim200-1000$\,s durations and integrated 3--79 keV fluxes in the range $10^{-12}$ to $10^{-10}$ \ergcms. Four of these are spectrally hard, with power-law indices $-3<\Gamma<0$, making them distinct compared to FXTs discovered in the soft band. With the exception of NuSTAR 230103 which has the softest spectrum of our FXT candidates, all sources presented here would have been too faint below 10 keV to be detected by EP, which has a sensitivity of $3\times10^{-11}$ \ergcms\ in the 0.5--4 keV band in a 1000\,s exposure assuming a $\Gamma=-2$ power-law \citep{yuan22}. Three have potential associations with galaxies at $z=0.1-2$, implying 3--79 keV luminosities of $10^{43}$ to $10^{48}$\ergs\ and volumetric event rates of 125--2900 Gpc$^{-3}$ yr$^{-1}$. The luminosity, duration and volumetric rates of our FXTs are consistent with low-luminosity GRBs, or in one case a supernova shock breakout, if these are indeed distinct phenomena. 
\facilities{NuSTAR} 

\software{{\tt HEASoft} \citep{heasoft}, {\tt NuSTARDAS}, {\tt XSPEC} \citep{arnaud96}}

\acknowledgements{We thank the anonymous referee for their insightful and valuable input into our manuscript, which improved its quality.

\nustar\ is a project led by the California Institute of Technology, managed by the Jet Propulsion Laboratory, and funded by the National Aeronautics and Space Administration. This research has made use of the NuSTAR Data Analysis Software ({\tt NuSTARDAS}) jointly developed by the ASI Science Data Center (ASDC, Italy) and the California Institute of Technology (USA).

This research has made use of data and/or software provided by the High Energy Astrophysics Science Archive Research Center (HEASARC), which is a service of the Astrophysics Science Division at NASA/GSFC.}

\bibliography{bibdesk.bib}

\end{document}